# Investigating Mental Representations about Robots in Preschool Children

**Monaco Camilla**

Provincial Federation of Preschools

Trento, TN-38123, Italy

camilla.monaco@fpsm.tn.it

**Ceol Tiziana**

Provincial Federation of Preschools

Trento, TN-38123, Italy

tiziana.ceol@fpsm.tn.it

**Mich Ornella**

Bruno Kessler Foundation

Trento, TN 38123, Italy

mich@fbk.eu

**Potrich Alessandra**

Bruno Kessler Foundation

Trento, TN 38123, Italy

potrich@fbk.eu

**Abstract**

This paper refers to an observational research that investigates preschool children's mental representations of robots. Our hypotheses were that: a) three to six years-old children think about robots as human-like entities, concerning to both the physical and the conceptual nature of human beings, and b) they do not understand the concept of the software programming behind the operation of a robot. The study is based on two different data collection systems: an individual-based system and a group-based one. In both cases, the investigation uses a complex and multimodal research structure that combines a drawing-based approach with a conversational-discursive methodology. This paper focuses on the individual-based data collection (348 drawings and 118 interviews). Preliminary results show that the human-like representation of robots is not the only one, even though it is present in about 64% of the drawings, and that this percentage decreases after the interview about robotics. Moreover, in several cases children refer to some concepts related to programming (e.g. the presence of sensors or human beings that make robots do things).

**ACM Classification Keywords**

H.5.m. Information interfaces and presentation (e.g., HCI): Miscellaneous.

**Introduction**

What are the modalities and strategies that children use to socially construct their mental representations [5, 8] about robots? In our perspective, children construct ideas, hypotheses and theories about the world they live in – and also about robotics – within and throug the significant social interactions that characterize their everyday life contexts (family, school, etc.) [7, 10, 11]. When we talk about educational robotics, we refer to all those educational activities that are based on designing, creating, implementing and operating with robots, i.e. machines that act following software programming. Educational contexts such as preschool can use robotics as an instrument to promote children's development and learning within a well-grounded educational planning framework [3, 4, 7, 9]. We think that robotics which is usually seen as particularly *difficult*, can characterize – and contribute to construct – educational contexts where knowledge is socially and collaboratively constructed by preschool children. In fact, throughout social interaction within significant and challenging situations, children can share competences and knowledge in order to co-construct original and complex ideas, representations and theories about robotics and about how it can be used at school [7, 11]. We know that children – even the youngest ones – produce *difficult* thoughts and are able to reason about complex systems. It seems that working to implement and make more comprehensive children's mental representations improves their learning and enhances their use strategies [1]. This paper refers to a large observational research that investigates preschool children's mental representations of robots.

Our hypotheses were that:

a) three to six years old children think about robots as human-like entities, concerning both physical and conceptual nature;

b) they do not have the concept of software programming behind the operation of a robot (i.e. they do not know that robots work because someone "tells them what to do").

The research was born to better comprehend what three to six years old children think about robots, within the specific socio-cultural context of the Italian Preschools, where robotics is still almost absent. On the one hand, we are interested to understand how robotics could be introduced as an instrument to support and promote children's learning processes. On the other hand, we are convinced that it is necessary to first discover and analyze children's complex ideas and theories concerning robots. Indeed, as Corsaro states [2], children are the best source to comprehend childhood

The study is based on two different data collection systems: an individual-based system and a group-based one [4]. In both cases the investigation uses a complex and multimodal research structure that combines a drawing-based approach with a conversational-discursive methodology.

In particular, this paper focuses on the individual-based data collection, that involves a corpus of 348 drawings with their audio-video recorded descriptions and 118 audio-video recorded interviews, collected at school by trained teachers. Concerning this part of the research, we decided not only to ask children to draw robots using their own ideas, but also to elicit their description of the drawings and to interview them about robotics.

Our idea was that drawing cannot be considered sufficient to elicit children's mental representations. We expected that the opportunity to explain their own drawing and, moreover, the possibility to discursively interact with the teacher about robotics would allow participants to give much more information about what they think [3, 6, 7].

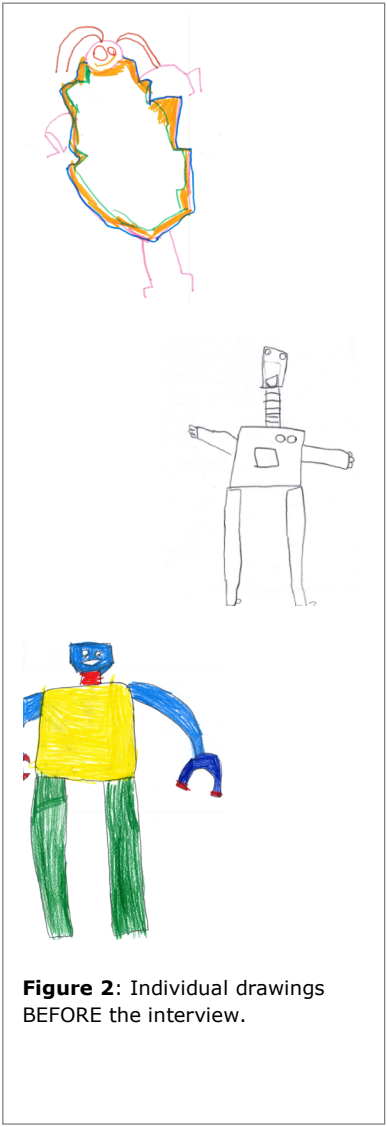

**Figure 2**: Individual drawings BEFORE the interview.

**Method**

The study consists of two different methodological steps:
1) individual-based data collection;
2) small group-based data collection. It is composed by two different groups of participants: a) *expert* children, i.e. those children who have participated in the individual-based data collection) and b) *beginner* children, who have not participated in the individual-based data collection.

*Participants*

219 preschool children (three to six years-old) and 25 teachers, belonging to twelve preschools associated with the Provincial Federation of Preschools of Trento, Italy.

*Instruments*

a) Individual-based research step (this step was video or audio recorded):
 - individual drawing;
 - individual description of the drawing;
 - semi-structured individual interview (based on some questions that were previously shared and discussed with the trained teachers). At the end of the interaction with the child, the teacher showed five pictures of robots asking "In your opinion, what are these?" (see Figure 1).

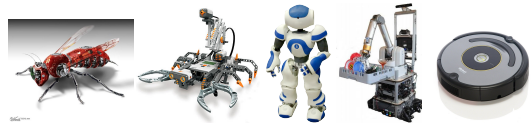

**Figure 1**. Images used for the individual interview.

b) Small group-based research step (this step was video or audio recorded):
 - collaborative drawing in a small group;
 - small group discussion;
 - small group meta-reflection about collaborative drawing.

*Procedure*

The individual-based research step, which is the focus of this paper, involved 121 children. Each participant took part in four different steps managed by a trained teacher:
a) first drawing of a robot, with the respective description;
b) semi-structured interview about robotics;
c) second drawing of a robot, with the respective description;
d) autobiographical drawing.

We collected the second drawing in order to investigate whether the conversational interaction (interview) about robotics contributes somehow to modify children's graphic productions. On one hand, since we know that children's mental representations are much more complex than their drawings, we decided to give them the opportunity to discursively express their opinions interacting with the teacher. Indeed, reasoning construction is not something individual and it can be strongly improved throughout significant social interaction [7, 10, 11]. On the other hand, we were interested to investigate whether after the discursive interaction with the teachers children graphically produce different forms of robots. In other words, we hypothesized that talking and reflecting about robotics, as well as watching some pictures of robots can also produce some modifications concerning children's drawing production. Moreover, the introduction of step d) was aimed to compare the way children usually draw their selves with their two graphic representations of robot. This is useful to understand if some features of the two robots drawn by the child are related to the usual characteristics of her/his drawn human beings. The trained teachers collected 348 drawings (119 first drawings, 115 second drawings, 114 autobiographical drawings) and 118 audio or video recorded interview.

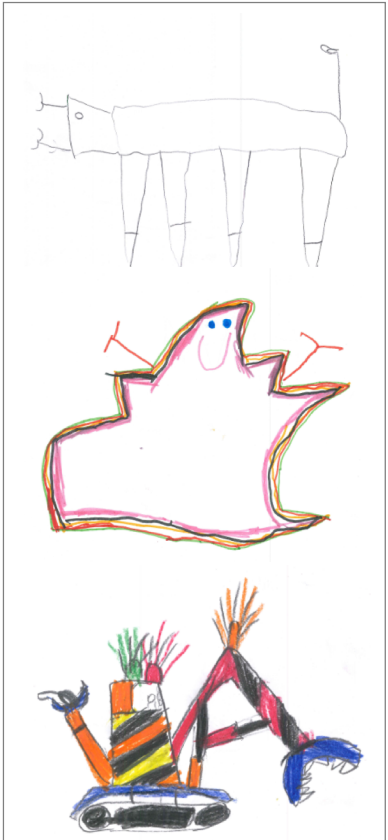

**Figure 3**: Individual drawings of the same children who drew the drawings shown in Figure 1, AFTER the interview.

**Preliminary results**

Concerning the initial hypotheses, preliminary results derived from the analysis of drawings show that:

- human-like representation of robots is not the only one present, even though it is present in about 64% of the first drawings (see few examples in Figure 2);
- human-like representation of robots tends to decrease after the interview about robotics: it is present in about 54% of the second drawings (see few examples in Figure 3).

Moreover, listening to some interviews and to descriptions of the drawings, we saw that some children mention some concepts related to programming (e.g. the presence of sensors or human beings that make the robot do things).

**Discussions and Future Work**

As reported in the previous section, we found some interesting preliminary answers to our starting hypotheses. However, a more systematic analysis of both the drawings and the interviews is now necessary to give statistical significance to our preliminary results. To this end, we are working on a validated method to categorize drawings. Moreover, we will analyze the video material previously annotated with a specific annotation tool, such as Anvil. We will also use the main principles of Conversational Analysis and Discourse Analysis to investigate the children-teachers social interaction. The preliminary analysis of the data we collected also gives interesting information on the methodology used. Indeed, the initial results confirm that in several cases the child's description and explanation of her/his drawing add very important information concerning her/his ideas and representations of robots. This additional information is particularly significant when the graphic performance of the child is not so clear and recognizable. Moreover, in most cases the individual interview highlights the richness and complexity of children's thinking processes about robotics.

**Acknowledgements**

We thank all the children, teachers and pedagogical coordinators who made this research possible and concrete.